\documentstyle[epsf]{mn}
\def\ion#1#2{{\rm #1}%
\ifmmode{\mathchoice{\scriptstyle}{\scriptstyle}
{\scriptscriptstyle}{\scriptscriptstyle}{\rm\uppercase{#2}}}%
\else$\,\scriptstyle\rm\uppercase{#2}$\fi}
\def\HII{{\ion{H}{II}}}
\def\refeq#1{{(\ref{#1})}}
\def\pdif#1#2{\mathchoice{\partial{#1}\over\partial{#2}}
{\partial{#1}/\partial{#2}}{\partial{#1}/\partial{#2}}
{\partial{#1}/\partial{#2}}}
\def\id{{\rm\,d}}                                
\def\etal{{et~al.}}
\def\eg{{e.g.}}
\def\ie{{i.e.}}

\def\cf{{cf.}}
\def\unit#1{\,{\rm {#1}}}
\def\umult#1#2{\ifx#2\unit\def\tmpa{\umulti#1#2}\else\toks0=\expandafter{#2}%
\edef\tmpa{\noexpand\umulti\noexpand#1\the\toks0}\fi\tmpa} 
\def\umulti#1#2#3{\ifx#2\unit\unit{#1#3}\else%
\message{\noexpand\umult error: must precede unit or \noexpand\unit}%
\unit{#1}#2#3\fi} 

\def\kms{\unit{km\,s^{-1}}}
\def\secnd{\unit{s}}
\def\yr{\unit{yr}}
\def\gram{\unit{g}}
\def\cm{\unit{cm}}

\edef\vbar{\char`|}                              
\def\subrm#1{_{\rm #1}}
\catcode`|=\active \let|=\subrm
\def\ee{\protect\pee}
\def\pee#1{\ifmmode{\times10^{#1}}\else$\times10^{#1}$\fi}
\newcommand{\UCHIIR}{{UCH{\sc ii}R}}
\newcommand{\domust}{{\dot{\mu}_\star}}
\newdimen \figwidth
\newif\ifref
\reffalse
\ifref
\figwidth \textwidth
\else
\fi
\title[Cometary mass loaded UCH{\small\it II}R]
{Clumpy ultracompact H{\Large\bf\,II} regions -- III.\\
Cometary morphologies around stationary stars}
\author[R.J.R. Williams, J.E. Dyson and M.P. Redman] 
{R.J.R. Williams$^1$, J.E. Dyson$^{1,2}$ and M.P. Redman$^1$\\
$^1$Department~of~Physics~and~Astronomy,
University~of~Manchester, Oxford~Road, Manchester M13~9PL\\
$^2$Department~of~Physics~and~Astronomy,
University~of~Leeds, Leeds LS2~9JT}

\date{Accepted 1996 February 12. Received 1996 January 29; 
in original form 1995 October 9}
\pagerange{\pageref{firstpage}--\pageref{lastpage}}

\begin{document}
\label{firstpage}
\maketitle

\begin{abstract}
Cometary ultracompact \HII\ regions have been modelled as the
interaction of the hypersonic wind from a moving star with the
molecular cloud which surrounds the star.  We here show that a similar
morphology can ensue even if the star is stationary with respect to
the cloud material.  We assume that the \HII\ region is within a
stellar wind bubble which is strongly mass loaded: the cometary shape
results from a gradient in the distribution of mass loading sources.
This model circumvents problems associated with the necessarily high
spatial velocities of stars in the moving star models.
\end{abstract}


\section{Introduction}

The ultracompact \HII\ regions (\UCHIIR) found deep within molecular
clouds provide important information on the early phases of the
interaction of massive stars with their natal environment.  The
disruption of the cloud material by the hypersonic winds and UV
radiation fields of these stars is a severe barrier to an
understanding of the process of massive star formation -- only by
studying the disruption process will anything be learnt about the
innermost regions of the protostellar cloud.  The disruption process
also involves many important problems of gas dynamics.  Considerable
theoretical effort has gone into modelling the varied morphology of
\UCHIIR\@.  

Most theoretical attention has so far been given to the cometary
regions, which comprise about 20 per cent of observed \UCHIIR\
\cite{chur90}.  Perhaps the most detailed model, that of Van Buren \&
Mac Low (1992, and references therein), treats them as the
steady-state partially ionized structures behind bow shocks driven by
the winds of stars moving through molecular cloud material.
Although it has been argued that some morphologies which are not
apparently cometary (in particular, core--halo) can be explained as
cometary structures viewed close to their axes \cite{maclea91}, the
shell and multiply peaked morphologies cannot, and so other models
should also be investigated.

There are also a number of unresolved questions with regard to the
cometary models.  First, the star is assumed to be moving through
relatively homogeneous molecular cloud gas.  Yet it is well known that
cloud material has a clumpy distribution down to very small scales.
Secondly, current cometary models require rather high stellar
velocities, characteristically 10--20$\kms$ (Van Buren \& Mac Low
1992; Garay, Lizano \& Gomez 1994).  The origin of such a high
velocity dispersion is problematic.  Moreover, as emphasized by Gaume,
Fey \& Claussen~\shortcite{gafc94}, if these high stellar velocities
are correct, the vast majority of \UCHIIR\ should be well isolated
from other structures for most of their lifetime.  Yet they often
appear closely associated with other radio continuum sources and
molecular cores \cite{galg94}.

Churchwell~\shortcite{chur95} argues that the parameters used by Van
Buren \& Mac Low~\shortcite{vbml92} have a fairly large range of
possible values.  In particular, the ambient density used in these
models is significantly smaller than has been observed ahead of some
\UCHIIR\ \cite{cesaea95}.  Since the shape of the bow shock is
determined by the ram pressure of the impinging molecular material,
such high densities will allow cometary regions to form for
substantially smaller relative velocities.  However, such high
densities are not common, and are likely to be confined to relatively
small clumps.  Since these clumps will not be immediately destroyed on
their passage through the bow shock, the model discussed in the
present paper may in fact be a more appropriate description of the
structure of these regions.

Finally, as more detailed observations become available, the
similarity of shape to the structures predicted by Van Buren \& Mac
Low \shortcite{vbml92} becomes less convincing.  Some of the cometary
\UCHIIR\ observed by Gaume \etal~\shortcite{gaumea95} have tails that
hook back around the centre of the nebula, while the bow shock model
predicts structures with bright arcs and diffuse tails -- Gaume \etal\
indeed propose that such arc-like regions be considered a separate
morphological class.  When observed at higher resolution, apparently
cometary structures often seem to dissolve into collections of
individual emission knots (\eg{} Kurtz, Churchwell \& Wood 1994),
rather than remaining as smooth structures.

Dyson~\shortcite{dyso94} and Lizano \& Cant\'o~\shortcite{lizaea95}
have suggested that the interaction of a massive star with clumpy
molecular material provides a natural alternative model for \UCHIIR,
which circumvents the problem of the relatively long ($10^5\yr$)
lifetimes required by their high frequency compared to field OB stars
\cite{chur90}.  Clumps act as localized reservoirs of gas which can be
injected into their surroundings by photoionization and/or
hydrodynamic ablation.  The ionized region is bounded by a
recombination front; that is to say, an \HII\ region is held at a
constant radius by the high gas density which results from the mass
loading.  The frictional heating of the flow in mass loading regions
may explain the very high temperatures inferred for some fraction of
the gas in the Sgr B2 F complex \cite{mehrea95}.

There are many possible variants of this basic premise.  Dyson,
Williams \& Redman \shortcite[1995 -- hereafter]{dywr95} have
described models in which a strong stellar wind mass loads from
clumps.  In the model presented in Paper I (as in the present paper),
the ionized gas flow is everywhere supersonic.  Redman, Williams \&
Dyson \shortcite[1996 -- hereafter]{rewd95} have discussed models
where either the wind is so weak or the mass loading is so great that
the interior flow shocks or is totally subsonic.

The models of Papers I and II are spherically symmetric, and reproduce
well many observed morphologies (\eg\ shell-like and
centre-brightened).  They cannot, of course, produce cometary
morphologies.  We here describe how a simple geometrical distribution
of mass loading centres can generate such morphologies.  This model
has two important properties.  First, it again exploits the clumpy
nature of clouds.  Secondly, the star can be at rest with respect to
the cloud.  The structures predicted are dependent only in detail on
the exact form of the mass loading distribution.  We consider here
only the simplest case, namely where the flow in the \UCHIIR\ remains
supersonic, and defer a discussion of the much more complex subsonic
flows, and those in which an initially supersonic wind goes through a
termination shock, to a later paper.

\section{Model}

We adapt the supersonic, dust-free, isothermal flow model of Paper I
to include a non-spherical distribution of mass loading sources.  As
in Paper I, we take the stellar wind momentum output rate, $\domust$,
as
\begin{equation}
\domust = 4\pi r^2 \rho v^2 =
{\it const}, \label{e:momm}
\end{equation}
where $r$ is the distance from the star, $\rho$ is the gas density and
$v$ is the velocity of the flow.  In this supersonic model, we assume
that the streamlines are essentially radial.  Hence, we assume
equation~\refeq{e:momm} holds within any solid angle element and take
the mass conservation equation to be
\begin{equation}
{1\over r^2}\pdif{}{r}\left(r^2 \rho v\right) 
	= \dot{q} = \dot{q}_0 \exp(r \cos\theta/\lambda), \label{e:load}
\end{equation}
where $\dot{q}$ is the mass loading rate and $\lambda$ is the scale
height of the mass loading distribution.  In this model, $\rho$ and
$v$ are thus functions both of the distance from the central ionizing
source, $r$, and of the azimuthal angle, $\theta$.

The distribution of mass loading has been chosen to vary as a function
of the distance, $x = r\cos\theta$, perpendicular to a certain plane
through the star (this is not the conventional correspondence, as the
Cartesian co-ordinates are oriented by the line of sight assumed to
the region, while the spherical polars are oriented by its symmetry
axis).  At large positive $x$, the mass loading is great, whereas it
decreases to a negligible level at large negative $x$.  Such a
distribution of mass loading might be expected where a young ionizing
star formed at the border of a molecular cloud.  Closer in to the
centre of the cloud, an increased density of mass loading clumps will
produce a higher mean mass loading rate; away from the cloud, the mean
density of material will decrease and the mass loading rate will drop.

Integrating equation~\refeq{e:load}, we find that, once the mass flux
is dominated by the mass loading material, the flow density is
\begin{equation}
\rho = {4\pi\over\domust}{\dot{q}_0^2\lambda^4\over \cos^4\theta} 
	{1\over\tau^2}\left[\left(\tau^2-2\tau+2\right)\exp(\tau)-2\right]^2,
	\label{e:rhosol}
\end{equation}
where $\tau = r\cos\theta/\lambda$.  The recombination front radius,
$R|R$ (which is a function of the angle, $\theta$), follows from the
Str\"omgren relationship in a solid angle element d$\Omega$:
\begin{equation}
\int_0^{R|R} \beta n^2 r^2\id{r}\id{\Omega} 
	= S_\star\left({\rm d}\Omega\over 4\pi\right),
\end{equation}
where $\beta$ is the case B recombination coefficient, $n =
\rho/\bar{m}$ is the number density of nucleons (we assume the gas
inside the recombination front is almost completely ionized), and
$S_\star$ is the total stellar output of Lyman continuum photons.
Hence
\begin{eqnarray}
\cos\theta &=& \lambda 
\left(64\pi^3\beta \dot{q}_0^4\over S_\star \domust^2\bar{m}^2\right)^{1/11}
I(\tau|R)^{1/11}\label{e:shell}\\
& = & \left(\lambda\over 8\ee{16}\cm\right) 
	\dot{q}_{-30}^{4/11} S_{48}^{-1/11} \dot\mu_{28}^{-2/11}
	I(\tau|R)^{1/11},\label{e:shellb}
\end{eqnarray}
where $\tau|R = R|R \cos\theta/\lambda$ and
\begin{equation}
I(\tau) \equiv \int_0^\tau 
	\left[\left(\tau^2-2\tau+2\right)\exp(\tau)-2\right]^4 
	{\id\tau\over \tau^2}.
\end{equation}
In equation~\refeq{e:shellb}, we have taken $\beta =
2\ee{-13}\cm^3\secnd^{-1}$, $\bar{m} = 2\ee{-24}\gram$, and defined
scaling variables $S_{48} \equiv S_\star/10^{48}\secnd^{-1}$,
$\dot\mu_{28} \equiv \domust/10^{28}\gram\cm\secnd^{-2}$ and
$\dot{q}_{-30} \equiv \dot{q}/10^{-30}\gram\cm^{-3}\secnd^{-1}$ (cf.\
Paper I).

\begin{figure}
\centering
\mbox{\epsfxsize=\figwidth\epsfbox[30 394 551 730]{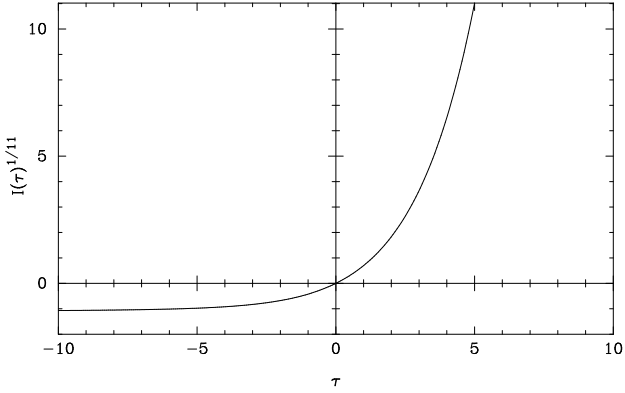}}
\caption{The function $I(\tau)^{1/11}$, which determines the shape
of the cometary regions (\cf\ equation~\protect\refeq{e:shell}).}
\label{f:itau}
\end{figure}
\begin{figure}
\centering
\mbox{\epsfxsize=\figwidth\epsfbox[38 438 539 686]{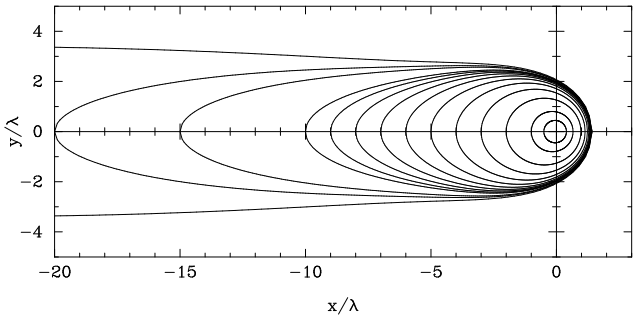}}
\caption{Cometary shapes, for a range of \UCHIIR\ sizes relative to the
scale height of the mass loading distribution.  The star is, in all
cases, at the intersection of the axes.}
\label{f:comets}
\end{figure}
\begin{figure*}
\centering
\begin{tabular}{cc}
\multicolumn{1}{l}{(a)} & 
\multicolumn{1}{l}{(b)} \\
\epsfysize=\textwidth\divide\epsfysize by 3
\mbox{\epsfbox[132 402 447 722]{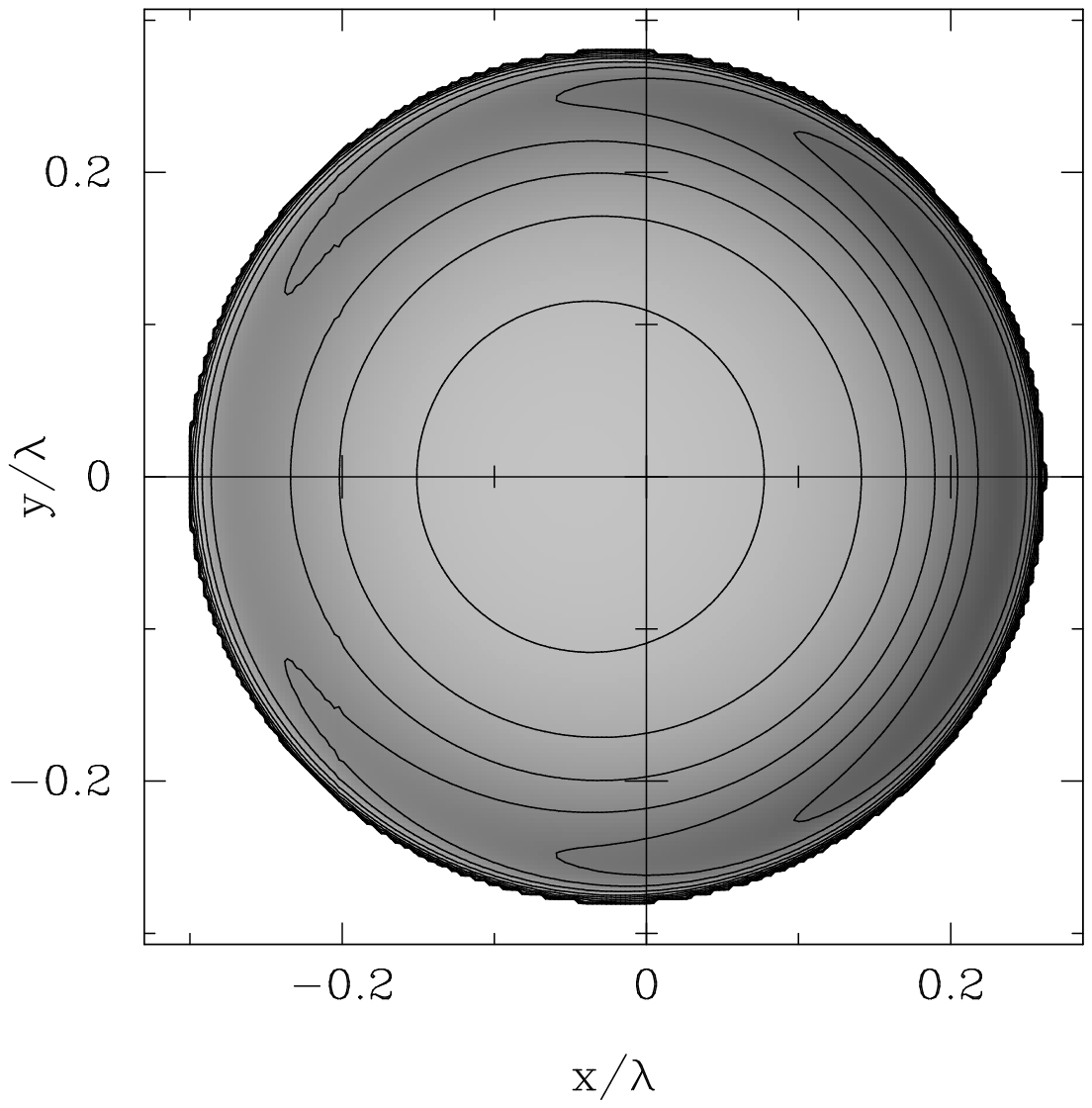}} &
\epsfysize=\textwidth\divide\epsfysize by 3
\mbox{\epsfbox[124 402 447 722]{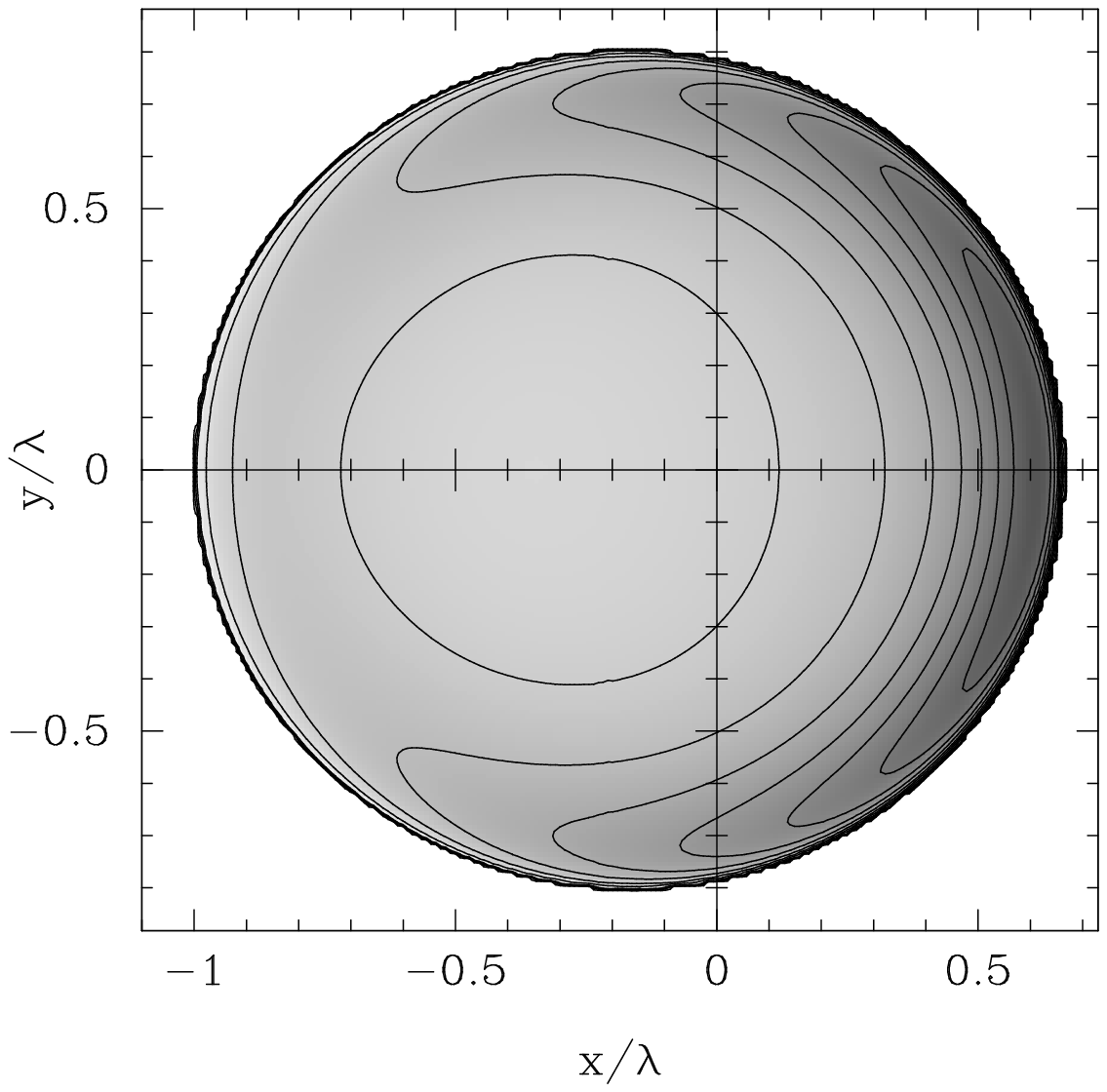}}
\end{tabular}
\caption{Grey-scale plot of emission from \UCHIIR\ with non-uniform
mass loading, with the recombination front in the direction of least
loading ($\theta = \pi$) at (a) $\tau|{tail} = -0.3$, (b) $\tau|{tail}
= -1$.  The star is at the intersection of the axes.  Contours are
also shown, at 2.5, 5, 7.5, 10, 20, 30, 40, 50, 60, 70, 80 and 90 per
cent of the peak intensity [the faintest contour away from the edge of
(a) is at 50 per cent, in (b) at 30 per cent].}
\label{f:comimage}
\end{figure*}
\begin{figure*}
\centering
\mbox{\epsfxsize=\textwidth\epsffile[37 414 560 722]{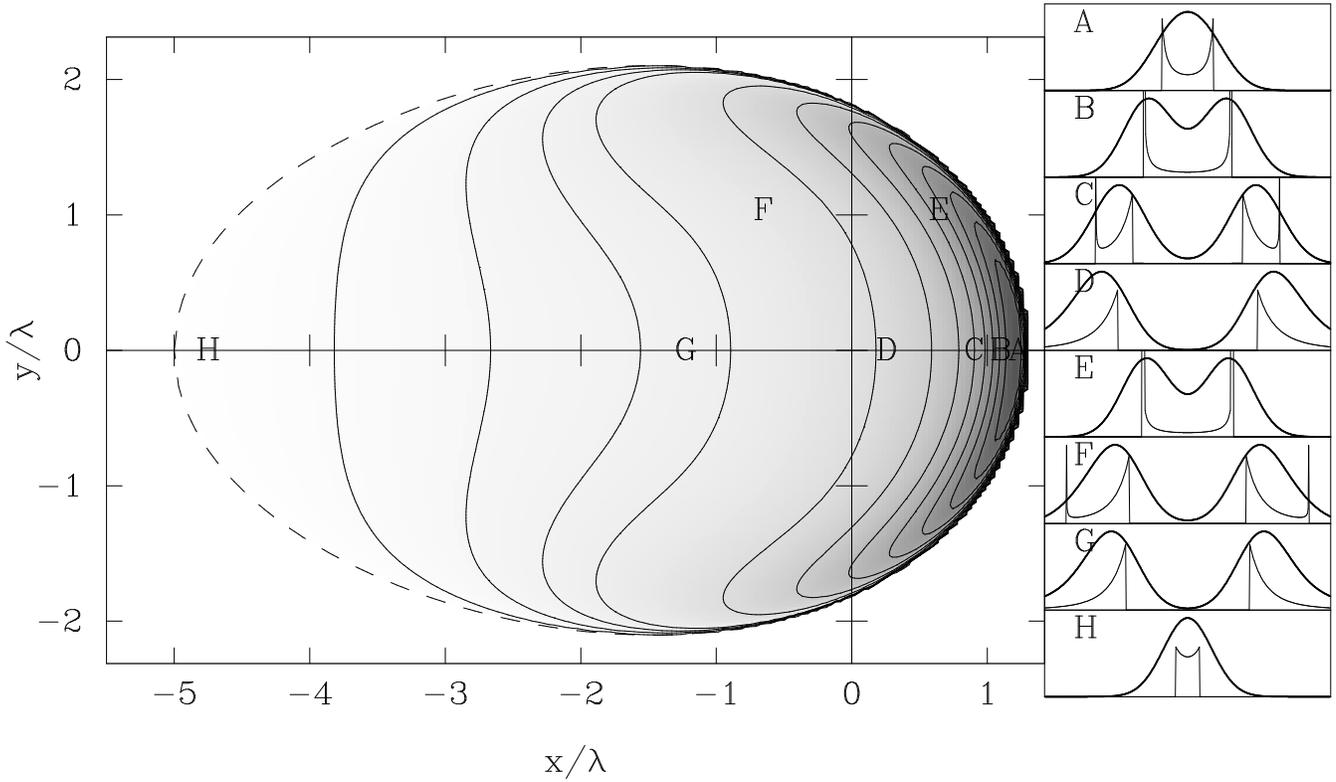}}
\caption{Grey-scale and contour emission measure image of a cometary
model, for $\tau=-5$ in the `tail'.  The contours are chosen as in
Fig.~\protect\ref{f:comimage}.  To the right, line profiles are shown,
calculated at the various labelled points.  The thin curves are the
zero-broadening line profiles, while the bold curves are shown for
thermal broadening appropriate to a radial Mach number of 2.2 at the
recombination front at the head of the cometary shape.  The velocity
scale is constant between the profiles, but the peak intensity scale
is arbitrary -- the total strength of the emission for each
observation may be judged from the emission measure value at that
point.}
\label{f:lineprofs}
\end{figure*}

In Fig.~\ref{f:itau}, we plot $I(\tau)^{1/11}$ as a function of
$\tau$.  From equation~\refeq{e:shell}, the recombination front
directly into and against the mass loading gradient occurs where
$I(\tau)=\pm{\it const}$: this will be symmetric if $\tau|R \ll 1$,
but, as $\tau|R$ increases, the ionized region will soon become highly
asymmetric.  The tail of the curve at large negative $\tau$ tends to a
constant value of $I(\tau)$: for high enough ionizing fluxes, the flow
in the direction of low mass loading is essentially free and the gas
remains ionized at all radii (within the assumptions of this model).

In Fig.~\ref{f:comets}, we plot the loci of $R|R(\theta)$ (\ie\ the
envelopes of the \UCHIIR\ when seen from a direction perpendicular to
the density gradient), for a range of \UCHIIR\ sizes relative to the
exponential scale height -- these curves are given parametrically by
equation~\refeq{e:shell}.

\subsection{Emission measure}

In this paper, we consider only the simplest case of cometary
structures observed edge-on.  We define Cartesian coordinates
$(x,y,z)$ with the star at the origin, where $(x,y)$ are in the plane
of the sky and $x=r\cos\theta$.  The emission measure at $x=b_1$,
$y=b_2$ is given by the integral along the line of sight (parallel to
$z$)
\begin{eqnarray}
{\cal E}(b_1,b_2) &=&\int n^2\id{z}\\
	&=& \dot{q}_0^4 \left(4\pi\over\domust\bar{m}\right)^2
	\left[(\tau^2-2\tau+2)\exp(\tau)-2\over \tau^3\right]^4 \nonumber\\ 
	& & \quad\times \int_{-z|R}^{z|R} r^8\id{z},
\end{eqnarray}
where $\tau=b_1/\lambda$ and $z|R$ is given by $z|R^2 =
b_1^2\tan^2\theta|R-b_2^2$, with $\theta|R$ given by
equation~\refeq{e:shell}.  

Example plots of the emission measure are shown in
Figs~\ref{f:comimage} and \ref{f:lineprofs}, with contours chosen to
be similar to those used in observational papers.  The size of the
cometary structures is parametrized by $\tau|{tail}$, the value of
$\tau|R$ in the direction of the tail, $\theta = \pi$.  The two cases
shown in Fig.~\ref{f:comimage} demonstrate the appearance of the
cometary form as the size of the \UCHIIR\ increases relative to the
scale of the non-uniformity in mass loading.  For $\tau|{tail} =
-0.3$, the region is mildly aspherical and may be compared to the
models of Paper I\@.  For $\tau|{tail} = -1$, the structure is similar
to that of the arc-like \UCHIIR\ described by Gaume
\etal~\shortcite{gaumea95}.  When $\tau|{tail} = -5$, the morphology
is cometary.

\subsection{Line profiles}

Optically thin line profiles can be calculated from the equation
\begin{equation}
{\cal E}(b_1,b_2,v|l) = \int_{-z|R}^{z|R} n^2f\left({z\over r} v_r 
	- v|l\right)\id{z},\label{e:lprof}
\end{equation}
where $(b_1,b_2)$ are the coordinates $(x,y)$ of the point observed in
the nebula, and the line-of-sight component of the velocity is $v|l$.
The distance $r$ is given, as above, as a function of $z$.  The flow
velocity, $v_r(b_1,b_2,z)$, is given by equations~\refeq{e:momm}
and~\refeq{e:rhosol} as
\begin{equation}
v_r = {1\over r^3} {\domust\over4\pi\dot{q}_0}
	\left[b_1^3\over(b_1^2-2b_1+2)\exp(b_1)-2\right],
\end{equation}
\ie, $v_r = A(b_1)/r^3$, where $A(b_1)$ is a constant for each
line of sight.  

In the limit of zero broadening (where $f(v) = \delta(v)$, the
Kronecker delta function), contributions to the integral,
equation~\refeq{e:lprof}, come at solutions, $z_i$, of the
quartic equation
\begin{equation}
(b^2+z_i^2)^2 v|l = z_iA(b_1), \label{e:critys}
\end{equation}
in the range $[-z|R,z|R]$, where $b^2 = b_1^2 + b_2^2$.  The line
profile is then given by
\begin{equation}
{\cal E}(b_1,b_2,v|l) = \sum_{\{z_i\}}
\left[n(b_1,b_2,z_i)\right]^2 
	{(b^2+z_i^2)^3 \over 
	A(b_1)\vbar b^2 - 3 z_i^2\vbar}.\label{e:deltaprof}
\end{equation}

Line profiles including broadening can be obtained by convolving this
line profile by a line-of-sight velocity function $g(v)$.  This
integration will remove the singularity in
equation~\refeq{e:deltaprof} where $z_i^2 = b^2/3$.  For instance,
thermal broadening is given by convolving with a Gaussian
\begin{equation}
g(v) = {1\over \sqrt{2\pi}\sigma} \exp(-v^2/2\sigma^2),\label{e:thermal}
\end{equation}
where the one-dimensional velocity dispersion, $\sigma$, is given by
$\sigma^2 = k|BT/M|Z = \bar{m}c^2/M|Z$ for line emission from a
species of atomic mass $M|Z$, if $\bar{m}$ is the mean mass per
particle and $c$ is the isothermal sound speed.  If the integral,
equation~\refeq{e:lprof}, is performed directly with $f(v)=g(v)$ given
by equation~\refeq{e:thermal}, major contributions to the total will
come from near the solutions $z_i$ of equation~\refeq{e:critys} and
from line-of-sight velocity extrema.

In Fig.~\ref{f:lineprofs}, we present line profiles for a selection of
points within the cometary shape.  Close to the edge of the region
(points A, H), the flow is nearly in the plane of the sky, and the
line widths are only slightly above the thermal widths.  For lines of
sight further into the nebula (points B, C, D), a larger component of
the flow velocity will be seen: the line profiles become increasingly
double-peaked.  Further around the bright shell (point E), the
double-peaked profiles are similar to those found along a line to the
apex.  The unbroadened profiles at points C and F show the sharp peak
at the maximum line-of-sight flow velocity, but when smoothed by the
thermal broadening this has little obvious effect.  At points D and G,
this spike is outside the range of velocities displayed in
Fig.~\ref{f:lineprofs}: the line profile has two fairly well-separated
components, with noticeably stronger tails to high velocity.

\begin{figure*}
\centering
\vbox to 14.5cm{\vfil
\mbox{\epsfxsize=\textwidth\epsffile[25 318 531 711]{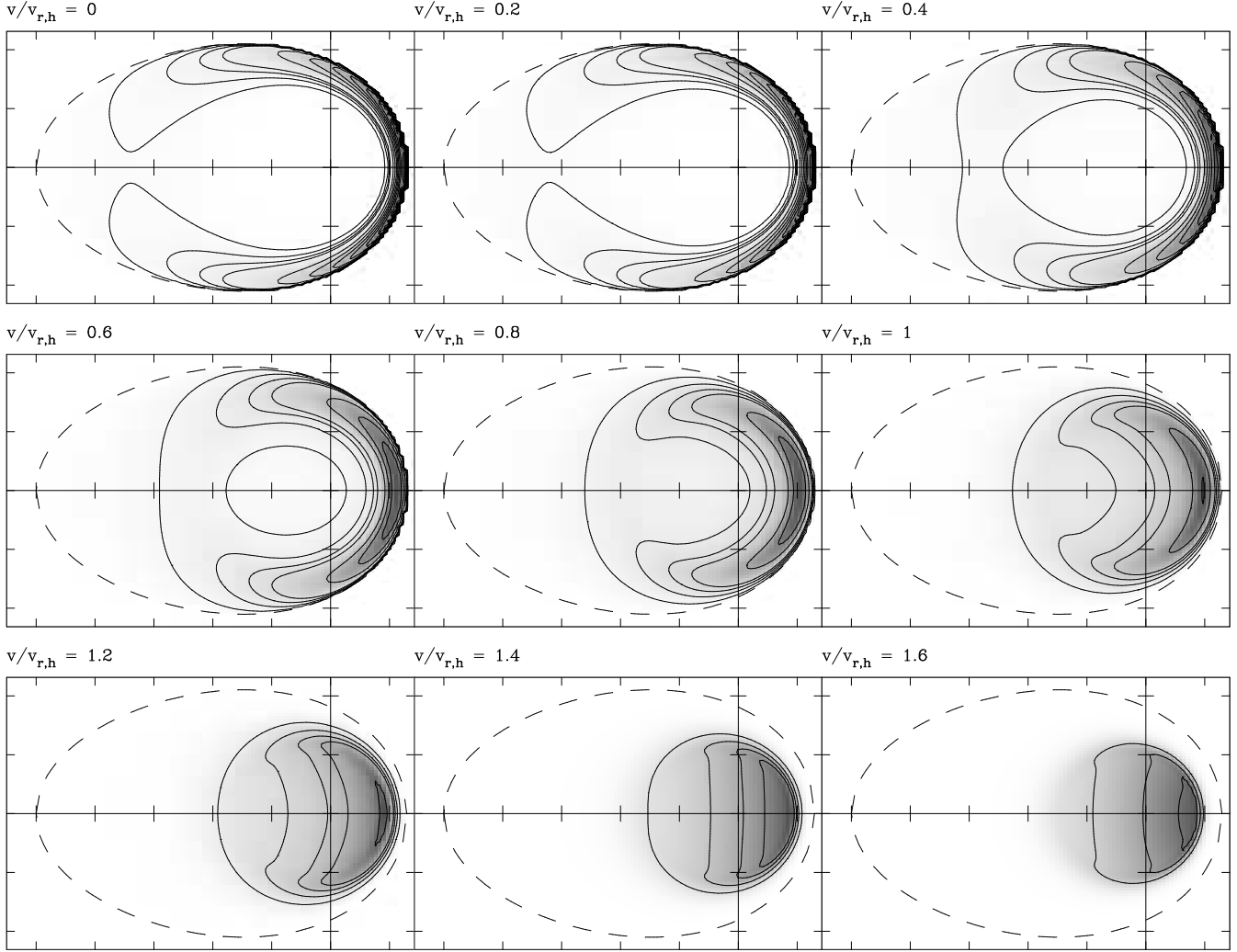}}
\vfil}
\caption{Channel maps for cometary \protect\UCHIIR, shown for velocities
$v$ as fractions of the radial velocity at the recombination front,
$v_{r,\rm h}$, at the head of the comet ($\theta=0$).  The profiles
are thermally broadened assuming that this velocity is $22\kms$, \ie\
Mach 2.2.  The grey-scale representation is normalized in each image;
the contours are scaled to 2.5, 5, 7.5, 10, 20, 30, 40, 50, 60, 70, 80
and 90 per cent of the peak intensity in the $v=0$ channel map.  The
ticks on the axis are at intervals of $\lambda$, the scale height of
the mass loading distribution; the dashed contour at the edge of the
region is as in Fig.~\protect\ref{f:lineprofs}.}
\label{f:channels}
\end{figure*}
In Fig.~\ref{f:channels}, we present the line profile data in the form
of maps at a single velocity (similar to observational `channel
maps').  These are shown for single velocities, $v$, but, since the
emission has been thermally broadened, they show the structures
expected for narrow-bandpass observations.  At low velocities, the
emission is bright, and is dominated by the dense gas at the edge of
the \UCHIIR\@.  As the relative velocity of the bandpass increases,
however, the emission peak gradually draws away from the edge of the
region and fills in the central hollow.

This displacement of the peak of the diffuse emission with increasing
relative velocity is a characteristic of radial flow models such as
that discussed here.  In contrast, in the bow shock and `champagne'
models, significant line-of-sight velocities can be produced at the
edge of the region by flows around the shell structure.  The finite
spatial resolution of radio observations and the presence of strong,
discrete emission components will, however, confuse this as an
observational diagnostic.  As can be seen from Fig.~\ref{f:channels},
the emission will only begin to become centre-brightened at relative
velocities $\ga 22\kms$, and with emission intensities $\la 10$ per cent
of the peak in the zero-velocity image.

These conclusions are, however, strongly dependent on the model
assumption that the dominant mechanism which broadens the emission is
thermal broadening.  As we argued in Paper I, the emission from
supersonic mass loaded winds is likely to be dominated by the dense
gas near to individual clumps.  The effect of the finite number of
individual clumps will be more noticeable in velocity-resolved
observations of \UCHIIR\@.  Also, the tails are likely to be very
dynamically active, with characteristic turbulent velocities similar
to the shear velocity between the wind flow and the (small) clump
velocity.  A dependence of this type would serve to fill in, at least
in part, the faint central regions of the low-velocity maps.  Firm
conclusions must, therefore, await a detailed treatment of the line
emission from the regions around individual mass loading clumps, which
is beyond the scope of this paper.

\section{Conclusions}

In this paper, we have shown how cometary morphologies may readily be
produced in a mass loaded model for \UCHIIR, with no need for relative
motion between the central young OB star and the molecular cloud
material, if mass loading sources are concentrated to one side.
Velocity channel maps show a shell structure stronger than the
integrated emission at small velocities relative to systemic, becoming
centre-brightened at higher relative velocities.  We have assumed that
the lines are broadened only to the thermal width: in reality,
however, the velocity structure may be dominated by turbulence in the
dense gas around individual mass loading sources.

We have assumed an exponential distribution in the rate of mass
loading with distance, $x$, away from a plane containing the star.
The general shape of the regions is not strongly dependent on the
detailed form of the mass loading distribution, so long as it has a
general decrease in one direction.  In one direction, the
recombination front is trapped close to the star by a high rate of
mass loading, while in the opposing direction the material remains
ionized to a rather larger radius, or even out to infinity.  Such
distributions might result not only near the edge of a molecular core,
but also if a small relative motion carried clumps into the \UCHIIR\
which were then evaporated as they crossed it.

Related models can be considered for different distributions of
mass loading centres: for instance, if the luminous star formed on a
ridge of clumpy molecular material (or surrounded by a thick disc),
the resultant \UCHIIR\ would have a bipolar morphology.  This can be
pictured by imagining a mirror placed at the plane $x=0$ of
Fig.~\ref{f:lineprofs}, to produce a structure of two back-to-back
cometary tails.

The model discussed here will give rise to observable velocity
gradients in the head-to-tail direction when the symmetry axis is
inclined to the plane of the sky.  Since the velocity of gas crossing
the recombination front, $v|R$, is nearly constant around the comet
(to within 50 per cent even when the tail is 50 scale lengths away
from the star), the maximum observed difference in mean line-of-sight
velocity will be $\sim 2v|R\sin i$, for inclination angle $i$.  Garay
\etal~\shortcite{galg94} find differences of $8$--$12\kms$ between the
mean velocities of emission in the head and tail of three cometary
\UCHIIR, which can be fitted with a mild inclination of the cometary
axis to the plane of the sky.

While Garay \etal~\shortcite{galg94} find no velocity gradients across
the regions they observe, Gaume \etal~\shortcite{gafc94} list three
well-defined cometary regions which have an asymmetric velocity
gradient perpendicular to the continuum symmetry axis of the
\UCHIIR\@.  This observation is problematic for the model discussed
here, as it is also for the bow shock model.  These contradictory
results may relate to morphology: Gaume \etal's sample tend to have
significantly longer and more complex tails, while Garay \etal's
sample are more arc-like.

We find that such arc-like morphologies are produced for milder
asymmetries in mass loading.  They form a natural intermediate between
shell and cometary morphologies.  Indeed, we would expect arc-like
morphologies also to be found in bow shock models for \UCHIIR, if the
relative velocity of cloud and star were low: such a model might,
however, have too low a gas pressure to correspond to observed
\UCHIIR\@.

Kurtz \etal\ \shortcite{kurcw94} find some observational evidence for
a size--density relationship for spherical and unresolved regions, but
not for cometary and core--halo \UCHIIR\@.  This suggests that a
Str\"omgren relation might hold for the former class, but not for the
latter.  However, given the sparse data on which these correlations
are based (which do not allow, for instance, for correction for the
luminosity function of the stars), we do not view this as a conclusive
argument against the model discussed in the current paper.

In subsequent papers, we will address the effects of tangential
pressure forces in mildly supersonic models and the structures
expected when the anisotropic mass loading is sufficiently strong to
produce a subsonic region of the flow.

\section*{Acknowledgments}

This work was supported by PPARC both through the Rolling Grant to the
Astronomy Group at Manchester (RJRW) and through a Graduate
Studentship (MPR).

\bsp
\label{lastpage}
\end{document}